\documentclass[reprint,amsmath,amssymb,aps,pra]{revtex4-2}
\usepackage{graphicx}
\usepackage{dcolumn}
\usepackage{bm}
\usepackage{amsmath}
\usepackage{array}
\usepackage{amsthm}
\usepackage{tabularx}
\usepackage{amssymb}
\usepackage{multirow} 
\usepackage{color}
\usepackage{float}
\usepackage{graphicx}
\usepackage{epstopdf}
\usepackage[caption=false]{subfig}
\usepackage[english]{babel} 
\usepackage{hyperref}
\usepackage[mathlines]{lineno}
\usepackage{mathtools}
\DeclarePairedDelimiter{\evdel}{\langle}{\rangle}
\newcommand{\av}{\evdel}

\newcommand{\change}[1]{ {\color{black} #1}}

\begin{document}
\renewcommand{\tablename}{Tabla} 
\renewcommand{\listtablename}{\'Indice de tablas}
\renewcommand{\figurename}{Figure}

\title{Markovian heat engine boosted by quantum coherence}

\author{Freddier Cuenca-Montenegro$^{1}$}
\author{Marcela Herrera$^{1}$} \author{John H. Reina$^{1,2}$}

\affiliation{$^{1}$Centre for Bioinformatics and Photonics—CIBioFi and Departamento de Física, Universidad del Valle, 760032 Cali, Colombia}
\affiliation{$^{2}$Department of Physics, 
University of Oslo, N-0316 Oslo, Norway}
\begin{abstract}
We evaluate the role of quantum coherence as a thermodynamic resource in a noisy, Markovian, one-qubit heat engine. 
By consuming the coherence of noisy quantum states, we demonstrate that the engine can surpass the classical efficiency limit when operating according to a quantum Otto cycle.
\change{The engine's non-classical nature is demonstrated by its violation of the Leggett-Garg's temporal correlations inequality. 
Amplitude damping increases the extractable work under partial thermalization, thereby increasing the efficiency.
In contrast, phase
damping increases the  extractable work under partial thermalization but reduces the efficiency}.
We implement the entire Otto cycle in a quantum circuit, simulating realistic amplitude and phase damping channels, as well as gate-level noise. We introduce an operational measure of the circuit’s thermodynamic cost to  establish a direct link between energy consumption and information processing in quantum heat engines.
\end{abstract}

\maketitle

\section{Introduction}
 
Quantum thermodynamics plays a pivotal role in the field of information theory,  accounting for the ergotropy of quantum systems~\cite{Binder2015,Campaioli2018,qbattery2025,francica2020,Yao2022,fox2024}, quantum coherence as a resource on the extractable work of a heat engine~\cite{Dillenschneider2009,Scully2003,Korzekwa2016,Camati2019, francica2020,Um2022,herrera2023,Aimet2023}, and the efficiency of a heat engine considering noisy states~\cite{Dionisis2014,miller2024power}, among others.  The realization that information is a physical resource akin to energy has led to groundbreaking advancements, such as quantum heat engines and quantum refrigerators. Through the integration of these breakthroughs, insight into the limits of energy manipulation and information processing has been acquired~\cite{qbattery2025,Goold2016}, thereby paving the way for progress in quantum computing, renewable energy, and other cutting-edge quantum technologies.

A fundamental tool for understanding thermodynamics in the quantum realm is the study of heat engines.
These engines are analogous to their classical counterparts
in that the working substance is replaced by a quantum system, such as spin-$\frac{1}{2}$ particles, which undergo transitions between the ground and excited states~\cite{Geva1992,Peterson2019, kumar2023,herrera2023}. This engine operates between hot and cold reservoirs and is limited to an efficiency below the classical limit~\cite{herrera2023,binder2018thermodynamics}. However, certain heat engines have been observed to utilize coherences or correlations as  quantum resources to enhance their efficiency and surpass the classical limit~\cite{herrera2023,binder2018thermodynamics,Niedenzu2018,Klatzow2019}.

Consequently, a body of research has emerged that explores the role of quantum coherences in the work extraction process within heat engines,  acknowledging the departure from classical principles~\cite{shi2020quantum}. 
These quantum coherences can be conceptualized as a measure quantifiable in terms of the relative entropy between the system's state and its dephased counterpart~\cite{francica2020, schlosshauer2019quantum}. This approach enables the study of physical quantities such as efficiency and work in open systems by observing the effect that coherence has on them. Consequently, various methods have been proposed to mitigate the loss of coherence in noisy states, with entanglement distillation being a notable example demonstrating the extractability of work from these states~\cite{miller2024power}.

Given the impact of quantum properties on the efficiency of a heat engine, it is relevant to study the quantum character of the system.   In this regard, the inequality proposed by Leggett and Garg~\cite{leggett1985quantum, emary2013leggett} holds for macroscopic systems but not for quantum systems. This inequality is a useful measure of non-classicality and allows for the detection of non-classical temporal correlations in the dynamics of the system~\cite{leggett1985quantum, emary2013leggett}.
In this paper, the extractable work and the efficiency of a single-qubit heat engine based on a quantum Otto cycle have been studied. This engine is based on an open system that allows one to observe the effect of noisy states on the measured observables and to verify the quantum properties of such an engine through the quantification of the Leggett-Garg time correlations and the coherence coming from the relative entropy.
\begin{figure}[b]
\centering 
\includegraphics[width=\linewidth]
{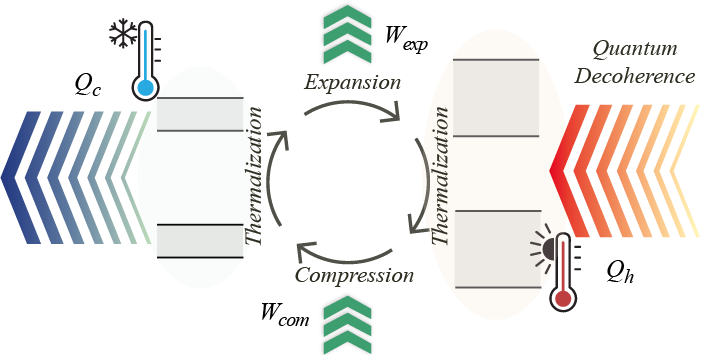}
\caption{Schematic of the quantum Otto cycle for the proposed one-qubit heat engine. The working substance is a spin-$\frac{1}{2}$ system and the cycle comprises two isochoric and two adiabatic strokes.  The cold and hot reservoirs are denoted by $c$ and $h$, respectively.}
\label{fig:Otto-cycle}
\end{figure}

\begin{figure*}
    \centering
\includegraphics[scale=0.3]{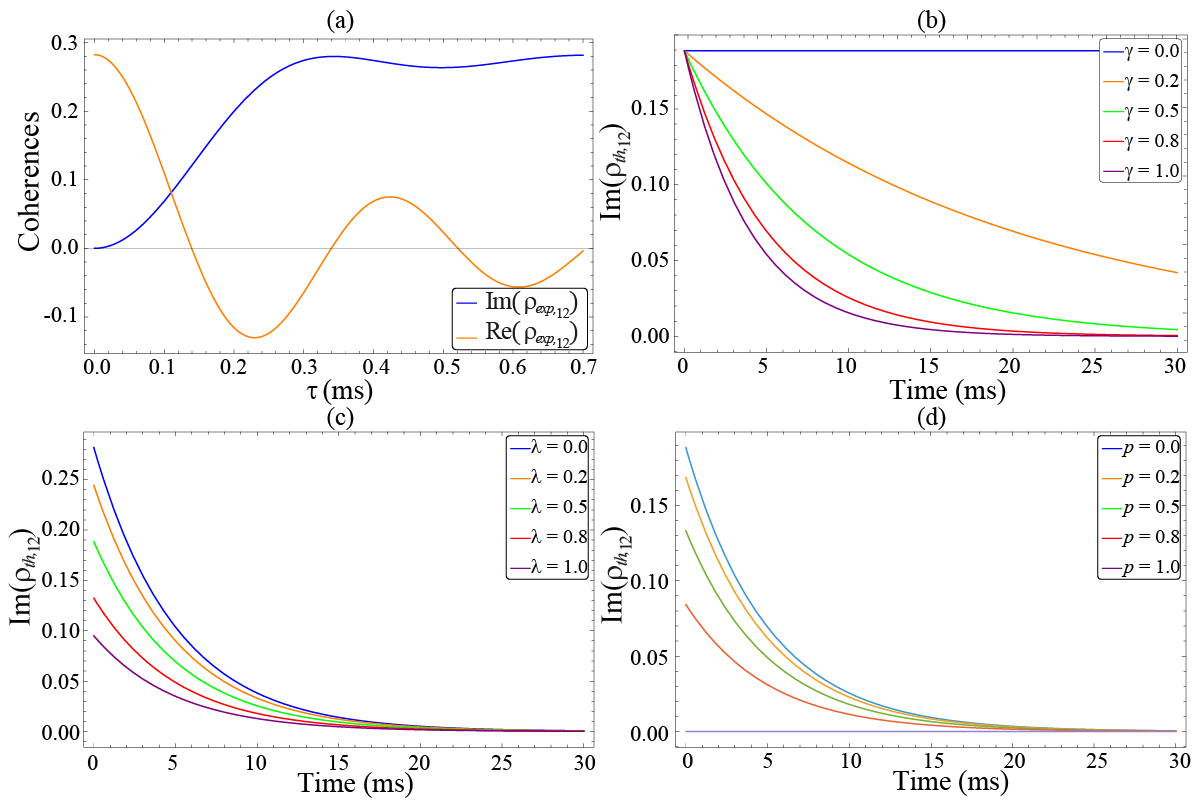}
\caption{\textcolor{black}{Quantum coherence dynamics of states  through the Otto cycle: (a) Production of quantum  coherence in the process of energy gap expansion carried out by the Hamiltonian $\mathcal{H}_{\rm exp}(t)$, Eq.~\eqref{Hexp}. (b) Decay of the imaginary part of quantum coherence in a thermal state $\rho_{\rm th}$, with a heat reservoir due to amplitude damping under partial thermalization $\lambda=0.5$ and $p=0$. (c)  Effect of thermalization on quantum coherence in the thermal state $\rho_{\rm th}$, with fixed amplitude damping $\gamma=0.8$. (d) Effect of phase damping on coherence of the thermal state $\rho_{\rm th}$, with $\lambda=0.5$ and $\gamma=0.8$. In all cases, the inverse temperatures and frequencies were $\beta_c=1.4$ (peV)$^{-1}$, $\beta_h=0.1$ (peV)$^{-1}$; $\omega_c=1.0$ rad/s, and  $\omega_h=1.8$ rad/s.}
}
\label{fig:Coherences}
\end{figure*}


\section{Model}

The one-qubit heat engine is based on a quantum Otto cycle, where the working substance is a spin-$\frac{1}{2}$ system. The cycle consists of two isochoric strokes and two adiabatic strokes, in the sense of quantum mechanics; that is, the process is slow enough to avoid transitions between energy levels.  \change{If $\epsilon_i$ and $\epsilon_j$ are the considered  Hamiltonian eigenvalues, then the  adiabatic time  $t_a>\pi/8(\epsilon_i-\epsilon_j)$}.

\change{The system begins in a thermal state with a compressed energy-level splitting. In this state, the ground and excited states are separated by an instantaneous energy gap, referred to as the “gap” throughout this work. This separation is determined by the difference in the Hamiltonian’s eigenvalues. The thermal state is  represented by the Gibbs distribution $\rho_c={\rm e}^{-\beta_c \mathcal{H}_c}/\mathcal{Z}_c$, with partition function $\mathcal{Z}_c=\mathrm{Tr}({\rm e}^{-\beta_c\mathcal{H}_c})$ and inverse temperature $\beta_c=(k_BT_c)^{-1}$ of a cold reservoir at temperature $T_c$, where $k_B$ is the Boltzmann constant.} \textcolor{black}{ $\mathcal{H}_c=\hbar\omega_c\sigma_x$ is the Hamiltonian of the system when it interacts with the cold reservoir of energy $\epsilon_c\equiv\hbar\omega_c$ through Pauli operator $\sigma_x$.  The cycle shown in Fig.~\ref{fig:Otto-cycle} consists of the following strokes.}

\begin{figure*}[ht]
    \centering  \includegraphics[width=1\linewidth]{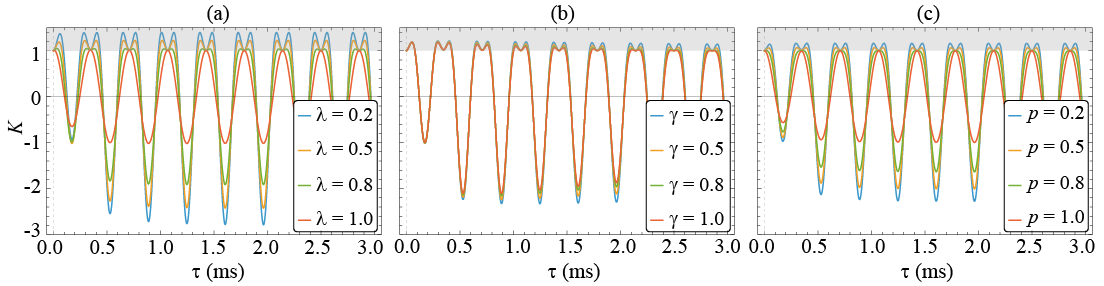}
    \caption{\change{Leggett-Garg time correlations for the Otto cycle in the heat engine. Temporal correlation function for (a) unitary evolution for different values of thermalization $\lambda$; $\gamma=0$, $p=0$,  (b) amplitude damping without dephasing ($p=0$),  $\lambda=0.5$,  (c) phase damping without amplitude damping ($\gamma=0$),  $\lambda=0.5$. Frequencies and inverse temperatures are the same as in Fig.~\ref{fig:Coherences}.}}
    \label{fig:correlations}
\end{figure*}

1. \textit {Expansion:} In this adiabatic stroke, the system has a \textit{gap} expansion described by the Hamiltonian,
\textcolor{black}{
\begin{equation}
    \label{Hexp}
\mathcal{H}_{\rm exp}(t)=-\hbar\omega(t)\left[\sigma_y\sin\left(\frac{\pi t}{2\tau}\right)+\sigma_x\cos\left(\frac{\pi t}{2\tau}\right)\right],
\end{equation}}
where \textcolor{black}{$\sigma_x$ and $\sigma_y$ are the usual Pauli operators},  
$\omega(t)=\omega_c+(\omega_h-\omega_c)\frac{t}{\tau}$ is the time-dependent frequency, \textcolor{black}{$\omega_c$ and $\omega_h$ are the angular frequencies of particles in the compressed and expanded states} ($\omega_c<\omega_h$), and $\tau$ is the operation time, i.e. the time taken to pass from a frequency $\omega_c$ to $\omega_h$, $0\leq t \leq \tau$.

2. \textit{Thermalization with hot reservoir:} The gap remains expanded, then this is an isochoric stroke. Here the system is coupled to a hot reservoir at temperature $T_h$, which transfers heat to the particles, causing some transitions from the ground state to the excited state~\cite{binder2018thermodynamics}. In this stroke we consider a partial SWAP thermalization, which can be represented by a dynamic mapping  $\rho_{\rm th}=\mathcal{E}_{\rm th}(\rho)=\sum_iE_i^{th}(\lambda)\rho_{\rm exp}E_i^{th}(\lambda)^\dagger$, where $\rho_{\rm exp}$ is the state of the system after the expansion process, such that for $\lambda=1$ ($\lambda=0$) there is complete (no)  thermalization \change{(see Appendix~\ref{ap_dynamicMapping})}. \textcolor{black}{$\mathcal{H}_h=\hbar\omega_h\sigma_y$} is the Hamiltonian defining the stroke; we set $\epsilon_h\equiv\hbar\omega_h$.

3. \textit{Compression:} After thermalization, the system is decoupled from the hot reservoir and starts a compression process, passing from a state with energy gap $\epsilon_h$ to one with $\epsilon_c$,
with a Hamiltonian 
\textcolor{black}{\begin{equation}
        \label{Hcom}
\mathcal{H}_{\rm comp}(t)=\hbar\omega(t)\left[\sigma_y\cos\left(\frac{\pi t}{2\tau}\right)+\sigma_x\sin\left(\frac{\pi t}{2\tau}\right)\right].
\end{equation}}
This considers that the time needed to compress the gap is the same time needed to expand it, $0\leq t \leq \tau$.
    
4. \textit{Thermalization with cold reservoir:} In this stroke, the system is coupled to a cold reservoir, at temperature $T_c$, and undergoes a process similar to the partial SWAP thermalization at the hot reservoir,  $\rho_{tc}\equiv\mathcal{E}_{\rm tc}(\rho)=\sum_iE_i^{tc}(\lambda_c)\rho_{\rm comp} E_i^{tc}(\lambda_c)^\dagger$, where $\rho_{\rm comp}$ is the state of the system after the compression process, and $E_i^{tc}\equiv E_i^{th}$ \change{(see Appendix \ref{ap_dynamicMapping})}. This last stroke is described by the Hamiltonian $\mathcal{H}_c$. \textcolor{black}{Although in this step, we allow the system to fully thermalize ($\lambda_c=1$), enabling the next cycle to start from the state $\rho_c$.}

To distinguish the effect of the dissipation channels on the thermodynamic cycle, we first performed the evolution for a closed system, i.e., 
the von Neumann time dynamics 
$\frac{\partial \rho_j}{\partial t}=-\frac{i}{\hbar}[\mathcal{H}_j,\rho_j]$, 
where index $j$ runs through all the states in the Otto cycle. The solution to this equation  is $\rho(\tau)=U(\tau)\rho U^\dagger(\tau)$, with the unitary time evolution operator $U(\tau)=\exp\left(-i/h\int_0^\tau \mathcal{H}dt\right)$.
We then considered  a system-environment interaction where states evolve according to a master equation within the Born-Markov approximation~\cite{carmichael2013statistical,mccauley2020accurate},  \change{$\frac{\partial \rho}{\partial t}=-\frac{i}{\hbar}[\mathcal{H},\rho]+\mathcal{D}_{ad}(\rho)$, 
\begin{equation}
\mathcal{D}_{ad}(\rho)=\sum_k[L_k\rho L_k^{\dagger}-\frac{1}{2}\{L_k^{\dagger}L_k,\rho\}].
\label{pmatrix}
\end{equation}}
Here, $L_k$ are the usual Lindblad operators. 
For a spin-$\frac{1}{2}$ system, $L=\sqrt{\gamma}\sigma_-\equiv (\sigma_x-i\sigma_y)/2$ and $L^{\dagger}=\sqrt{\gamma}\sigma_+\equiv (\sigma_x+i\sigma_y)/2$, with a coupling parameter $\gamma$, raising and lowering operators $\sigma_+$ and $\sigma_-$, and Pauli operators $\sigma_x$ and $\sigma_y$. 
The solution of the master equation 
with the dissipator~Eq.~\eqref{pmatrix}, considering that the reservoir acts as a perfect heat bath~\cite{colla2024thermodynamic}, leads to a state that can be equivalently described by an \textit{amplitude} damping channel \change{(See Appendix \ref{ap_masterEQ})}, 
$\rho(t)=\mathcal{E}_{\rm ad}(\rho)=\sum_iE_i^a(t)\rho E_i^a(t)^\dagger$,
for a coupling parameter $\gamma$ representing the transition rate, $\gamma'=1-{\rm e}^{\frac{-\gamma}{2} t}$, and Kraus operators $E_i$~\cite{breuer2002theory,nielsen2001quantum}
\begin{equation}
    E_0^a=
    \begin{pmatrix}
        1 & 0 \\
        0 & \sqrt{1-\gamma'}
    \end{pmatrix} \hspace{1cm}
        E_1^a=
    \begin{pmatrix}
        0 & \sqrt{\gamma'}\\
        0 & 0 
    \end{pmatrix}.
\end{equation}
In addition, a \textit{phase}  damping channel is implemented \change{through the dissipator $\mathcal{D}_{pd}(\rho)=\Gamma(\sigma_z\rho\sigma_z-\{\sigma_z^\dagger\sigma_z; \rho\})$, where $\Gamma$ is the \textit{phase} damping rate, to obtain the dynamical mapping} $\rho=\mathcal{E}_{\rm pd}(\rho)=\sum_iE_i^p\rho E_i^{p\dagger}$, with Kraus operators 
\begin{equation}
    \label{eq:krausPD}
    E_0^p=\begin{pmatrix}
        1 & 0\\
        0 & \sqrt{1-p}
    \end{pmatrix} \hspace{1cm}
    E_1^p=\begin{pmatrix}
        0 & 0\\
        0 & \sqrt{p}   
    \end{pmatrix},
\end{equation}
and the decoherence parameter $0\leq p\leq1$. \change{See Appendix~\ref{ap_masterEQ} for details}. 

\begin{figure*}[t]
\centering
\includegraphics[width=\linewidth]{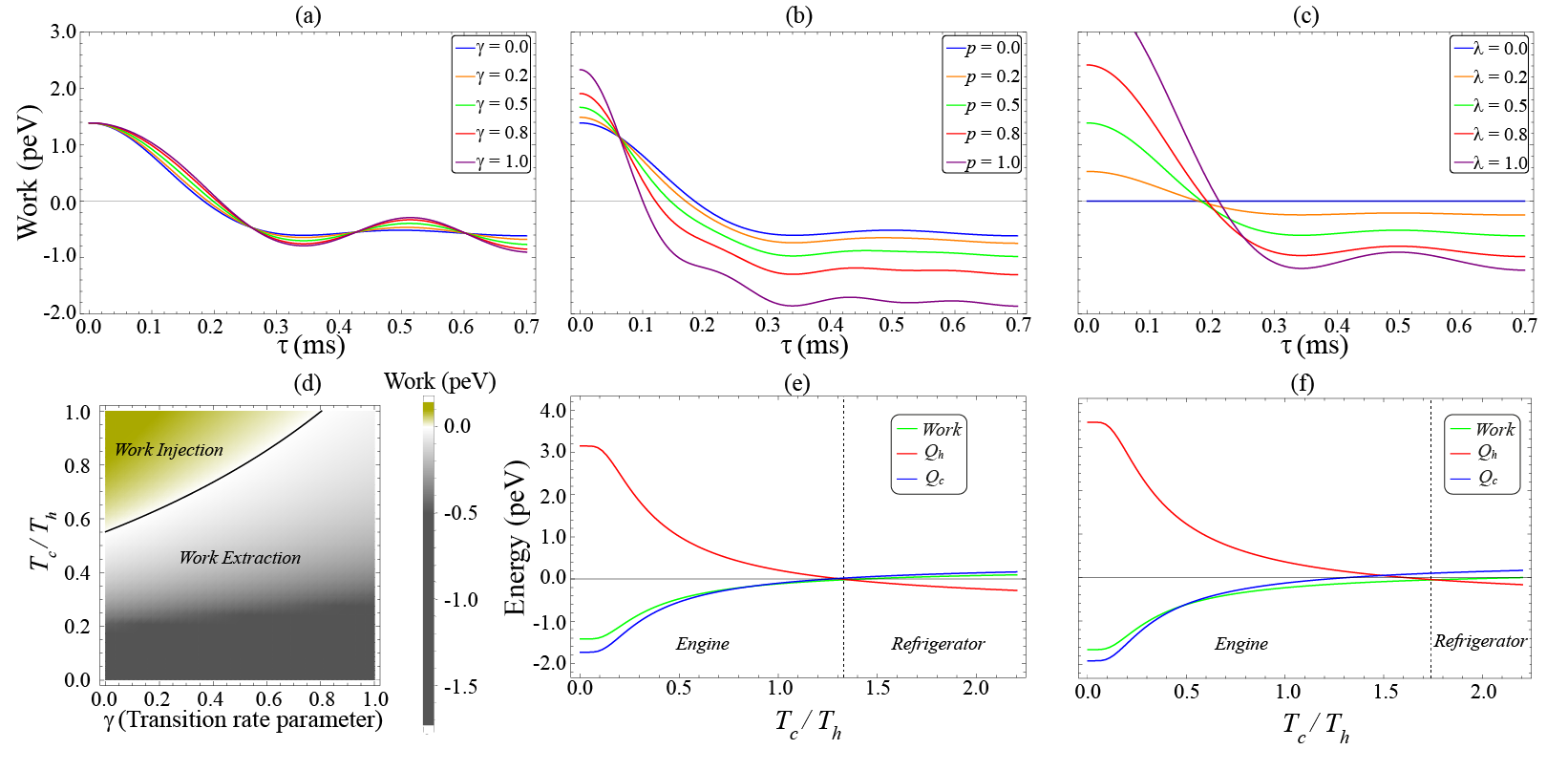}
\caption{\textcolor{black}{(a) Effect of amplitude damping $\gamma$ on the work dynamics with $p=0$, (b) Effect of phase damping $p$, without amplitude damping. (c) Effect of thermalization $\lambda$, for  $p=0$ and $\gamma=0$. (d) Phase diagram of the total work parameters for the heat engine,  $p=0$}, (e) and (f) operation regime of the heat engine and its variation with respect to power dissipation by amplitude damping; \textcolor{black}{$\gamma=0.2$ and $\gamma=0.8$, respectively. In graphs (a-c), $T_c=0.075\;\mu$K, $T_h=0.45\;\mu$K.  In all the graphs,  $\omega_c$ and $\omega_h$ are the same as in Fig.~\ref{fig:Coherences}$, \lambda=0.5$ (except in (c)).}
}
\label{fig:RegimeHeatEngine}
\end{figure*}

\section{Temporal correlations}

We demonstrate the quantum nature of the model by analyzing the Leggett-Garg time correlations~\cite{leggett1985quantum,emary2013leggett}. Here, for an observable $\mathcal{O}$ with two possible eigenvalues, it is possible to measure the correlations at two different times $t_m$ and $t_k$ by $C_{k,m}=\left< \mathcal{O}(t_k) \mathcal{O}(t_m)\right>$. For a macroscopic system, the following inequality holds~\cite{emary2013leggett}
\begin{equation}
    K\equiv C_{1,2}+C_{2,3}-C_{1,3}<1.
    \label{Leggett-Garg}
\end{equation}

\change{In our analysis, we use violation of the Leggett-Garg inequality (LGI), Eq.~\eqref{Leggett-Garg}, as an operational witness of quantum coherence~\cite{emary2013leggett}.  LGI violation implies that the system cannot be described by a classical stochastic process involving  non-invasive measurements. This nonclassical behavior is associated with the presence of quantum coherence in the energy eigenbasis and  enables coherent superpositions to persist across different stages of the Otto cycle}. \change{Here, we} choose the observable \change{$\mathcal{O}\equiv\sigma_z$} and measure it at three different times: the beginning of the gap expansion $t_1$, the end of the gap expansion $t_2$, and the end of the gap compression $t_3$,  $t_1<t_2<t_3$.


\change{As demonstrated in Fig.~\ref{fig:correlations}(a),  
 the observed LGI violation evidences the quantum nature of the proposed heat engine and establishes that its temporal correlations are non-classical.
 This violation is intrinsically linked to the presence of coherences formed during the energy gap expansion stage, as shown in Fig. \ref{fig:Coherences}(a), which persist during the partial thermalization ($\lambda < 1$). Maximum violation occurs at small values of $\lambda$, when the system retains significant coherence from its initial state while partially engaging with the hot bath (see Fig. \ref{fig:Coherences}(c)). This result  shows that the engine's quantum advantage depends on operating in a regime that preserves this coherence. As the thermalization parameter $\lambda$ approaches unity (complete thermalization), the system evolves into a fully thermal, diagonal state. This  leads to the loss of coherences and, consequently, to correlations $K \le 1$.} 

\change{
Furthermore, the quantum nature of the engine makes it  highly susceptible to environmental noise. As Fig.~\ref{fig:correlations}(b) shows, introducing amplitude damping models energy relaxation, which causes an exponential decay in the correlation function $K$ over time. This effect  becomes more pronounced with an increasing transition rate $\gamma$. Similarly, Fig.~\ref{fig:correlations}(c) shows that phase damping causes global decay of the temporal correlations as the parameter $p$ increases. Both 
channels rapidly suppress quantum coherences, as shown in Figs.~\ref{fig:Coherences}(b) and \ref{fig:Coherences}(d), driving the system toward classical temporal dynamics.  Appendix~\ref{ap_sigmaZ} presents a special case in which the Hamiltonians in the $\sigma_z$ basis allow us to obtain an analytical solution for $K$ (See Eq.~\eqref{eq_correlations}). This solution illustrates the influence of the parameters $p$, $\gamma$, and $\lambda$.}
\begin{figure*}[t]
    \centering
\includegraphics[width=1.0\linewidth]{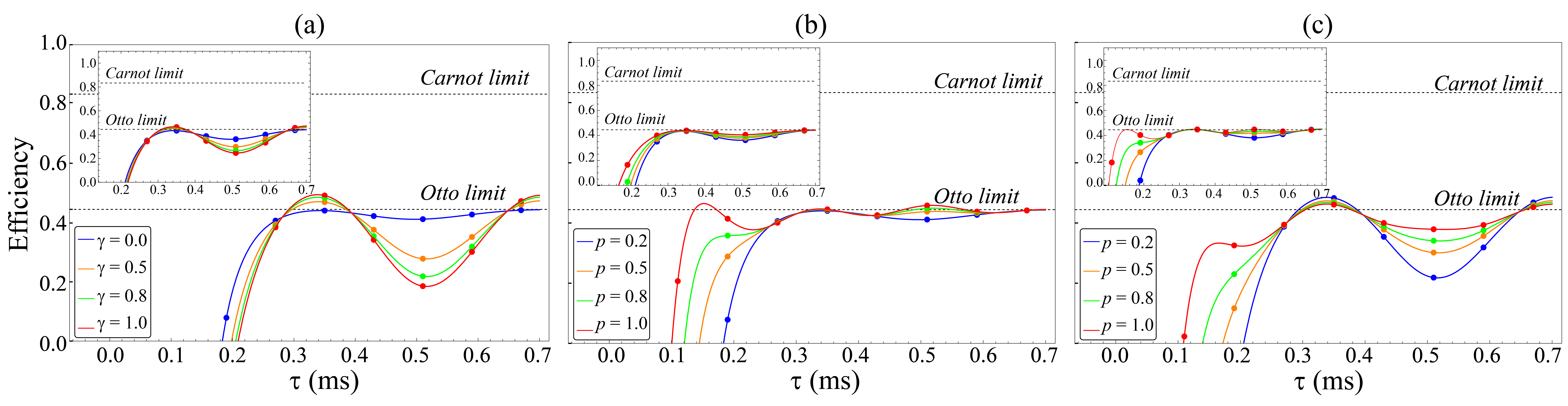}
\caption{\textcolor{black}{Effect of thermalization and noise on the efficiency of the heat engine: (a) Effect of amplitude damping $\gamma$ on the efficiency  for partial thermalization with the hot reservoir, $\lambda=0.5$ and $p=0$. The inset shows a similar situation with full thermalization, $\lambda=1$. (b) Effect of phase damping $p$, 
with $\gamma=0$ and $\lambda=0.5$; the inset shows this effect for $\lambda=1$. (c) Combined effects of amplitude and phase damping,  $\gamma=0.8$ (main figure),  and $\gamma=0.1$ (inset); $\lambda=0.5$.
In all cases, the inverse temperatures and frequencies were $\beta_c=1.4$ (peV)$^{-1}$, $\beta_h=0.1$ (peV)$^{-1}$; $\omega_c$ and $\omega_h$ are the same as in Fig.~\ref{fig:Coherences}. Continuous curves correspond to the efficiency obtained from Eq.~\eqref{eq_eta}, while the dotted curves correspond to the efficiency obtained with Eq.~\eqref{eq_efficiency}.}
}
\label{fig:efficiency}
\end{figure*}

\section{Extractable Work and Heat}
 
If the final states of the expansion (exp) and compression (comp) processes are given by \change{$\rho_{\rm exp}(\tau)=\exp\left(-i/h\int_0^\tau\mathcal{H}_{\rm exp}dt\right)\rho_c\exp\left(i/h\int_0^\tau\mathcal{H}_{\rm exp}dt\right)$, $\rho_{\rm comp}(\tau)=\exp\left(-i/h\int_0^\tau\mathcal{H}_{\rm comp}dt\right)\rho_{th}^f\exp\left(i/h\int_0^\tau\mathcal{H}_{\rm comp}dt\right)$} and the final state after partial thermalization is  $\rho_{\rm th}^f=\mathcal{E}_{\rm th}(\mathcal{E}_{\rm ad}(\mathcal{E}_{\rm pd}(\rho_c)))$, then, the work associated with each stroke is found to be the change in energy at each instant:
\begin{align}
\label{eq_work_exp}
    \av{W_{\rm exp}} = \; & \mathrm{Tr}[\rho_{\rm exp}(\tau)\mathcal{H}_{\rm exp}(\tau)]-\mathrm{Tr}[\rho_c\mathcal{H}_c],\\
\label{eq_work_comp}
    \av{W_{\rm comp}} = \; & \mathrm{Tr}[\rho_{\rm comp}(\tau)\mathcal{H}_{\rm comp}(\tau)]-\mathrm{Tr}[\rho_{\rm th}^f\mathcal{H}_h],
\end{align}
and the total work is the sum of the expectation values found in these processes:
\begin{equation}
    \av{W}=\av{W_{\rm exp}}+\av{W_{\rm comp}}.
\end{equation}

\change{The average work, as defined by Eqs.~\eqref{eq_work_exp} and \eqref{eq_work_comp} for expansion and compression processes, respectively,  is the difference in energy between the final and initial states of each process. Therefore, negative work values imply that the system is doing work on the environment. Conversely, positive work values indicate that the environment performs work on the system, i.e., the negative values of $\av{W}$ are the extractable work of the heat engine. During expansion and compression, the von Neumann entropy of the system remains constant. Thus, the energy change during these processes is due only to work.}

\change{Similarly,} the expectation values of the heat injected into and extracted from the system can be found as a change in the internal energy of the system in the thermalization strokes with the reservoirs,
\begin{align}
    \av{Q_h}=\; &\mathrm{Tr}[\rho_{\rm th}\mathcal{H}_h]-\mathrm{Tr}[\rho_{\rm exp}(\tau)\mathcal{H}_h],\\
    \av{Q_c}=\; &\mathrm{Tr}[\rho_{\rm c}\mathcal{H}_c]-\mathrm{Tr}[\rho_{\rm comp}(\tau)\mathcal{H}_c],
\end{align}
where $\av{Q_h}$ is the heat injected from the hot reservoir and $\av{Q_c}$ is the heat extracted from the cold reservoir.

\textcolor{black}{The results presented in Figure \ref{fig:RegimeHeatEngine}(a) demonstrate that the total work extracted from the quantum heat engine (QHE) increases with the parameter $\gamma$, which models the amplitude damping during the hot thermalization stroke. This behavior is specific to the partial thermalization regime (where $\lambda<1$), where the system remains far from thermal equilibrium with the hot bath during the process~\cite{Mohan_2025}. In this finite-time regime, increasing the damping parameter $\gamma$ effectively speeds up the thermalization rate relative to the fixed cycle time~\cite{dann2020quantum}. Consequently, a larger $\gamma$ drives the system closer to the fully thermalized state (as shown in Fig. \ref{fig:Coherences}(b)) that would be reached with a much longer interaction time. Since maximum energy is absorbed (and thus maximum work is available for extraction) when the system approaches thermal equilibrium with the hot reservoir, faster thermalization enhances work extraction by maximizing the internal energy of the working substance before the compression stroke.}

\change{
Figures~\ref{fig:RegimeHeatEngine}(b-c) show that full thermalization with the heat reservoir ($\lambda=1$) and increased phase damping ($p=1$) maximizes the extractable work in the QHE. These processes cause quantum decoherence, as shown in Figs.~\ref{fig:Coherences}(c-d), bringing the system to a state similar to complete thermalization, in which more work can be extracted.}

Figure~\ref{fig:RegimeHeatEngine}(d) shows two main zones. One zone for work extraction,  \change{$\lambda=0.5$}, is generated by setting small temperature ratios in the reservoirs. Larger amounts of extracted work are reached when the amplitude damping parameter is \change{increased}. The other zone corresponds to injected work and  increases with high reservoir temperature ratios and \change{small amplitude damping values,} $\gamma$. Combined with the fact that partial thermalization \change{maintains quantum} coherences~\cite{sutantyo2024performance} and reduces the amount of extractable work (see Fig.~\ref{fig:RegimeHeatEngine}(c)), this result 
 shows that \change{in the} system-environment interaction, less work can be extracted when more coherence is \change{present in the system's state}.

\change{\subsection{Operation regimes}}

The purpose of the proposed model is to use the quantum Otto cycle as a heat engine, so we analyze the conditions under which the quantum system can operate in this way. We use the following regime definitions~\cite{herrera2023}: 
\change{i) \textit{Heat engine:} 
$\av{Q_h}>0, \av{W}<0, \av{Q_c}<0$, and ii) \textit{Refrigerator:}
$\av{Q_h}<0, \av{W}>0, \av{Q_c}>0$.}
%
\change{In Figs.~\ref{fig:RegimeHeatEngine}(e-f)}, it can be seen that the described model has \change{two} modes of operation: the heat engine, which converts the heat injected from a reservoir into useful work, and 
the refrigerator, which removes heat from the system. 
\change{Thus, an increase in dissipation due to amplitude damping results in an increase in the regime in which the quantum system acts as a heat engine, while the refrigerator regime decreases}.
 
 
\section{Efficiency and Coherence}

\textcolor{black}{
We introduce quantum coherence into the states by applying Hamiltonians $\mathcal{H}_{\rm exp}(t)$ and $\mathcal{H}_{\rm comp}(t)$, (see equations \eqref{Hexp} and \eqref{Hcom}) in the rotating frame, as well as the Pauli operators $\sigma_x$ and $\sigma_y$. Quantum coherence is produced during the expansion process, as shown in Fig.~\ref{fig:Coherences}(a). However, once the system comes into contact with the heat reservoir, quantum coherence decays exponentially due to the combined effects of thermalization, phase damping, and amplitude damping, as illustrated in Figs.~\ref{fig:Coherences}(b-d). These conditions leave the working substance in a non-equilibrium state, with efficiency  calculated as
\begin{equation}
\label{eq_eta}
    \eta=-\frac{\left<W\right>}{\left<Q_h\right>},
\end{equation}
where $-\left<W\right>$ is the extracted work and $\left<Q_h\right>$ is the injected heat. The non-equilibrium state created by the noise effects and partial thermalization at this stage of the cycle results in a more favorable ratio of work to injected heat in Eq.~\eqref{eq_eta}, thereby increasing the efficiency beyond Otto's limit, $\eta=1-\frac{\nu_c}{\nu_h}$, where  $\nu_i=\omega_i/2\pi$, $i=c,h$~\cite{Dorfman_2018, Mohan_2025}.
}

\textcolor{black}{The interaction of the system with the reservoirs has associated entropy production and can be  related to the efficiency of the heat engine as follows:}
\textcolor{black}{
\begin{equation}
\label{eq_efficiency} \eta_E\equiv\eta_{Carnot}-\frac{\Delta S_\text{th}}{\beta_c\av{Q_h}},
\end{equation}
}
\textcolor{black}{where $\eta_{Carnot}=1-\frac{\beta_h}{\beta_c}$ is the Carnot efficiency, $\Delta S_{th}=S(\rho||\sigma)-\mathrm{Tr} \left[ \left( \rho_\text{th}^f - \rho_\text{exp}(\tau) \right)\ln \frac{\rho_\text{th}^f}{\rho_h} \right]$ contains all the entropy contributions when the system is coupled to the heat reservoir. $\rho_h=\mathrm{e}^{-\beta_h\mathcal{H}_h}/\mathcal{Z}_h$ is the fully thermalized state with the heat reservoir,  $S(\rho||\sigma)=D\left[\rho_{\rm exp}(\tau)||\rho_{\rm th}^f\right]+D\left[\rho_{\rm comp}(\tau)||\rho_{c}^f\right]$ gives the total relative entropy, and the quantum relative entropy 
$D\left[\rho||\sigma\right]=\mathrm{Tr}(\rho \ln\rho)-\mathrm{Tr}(\rho \ln\sigma)$ quantifies the distinguishability between the states $\rho$ and $\sigma$.
It can be shown that Eq.~\eqref{eq_efficiency}  is equivalent to Eq.~\eqref{eq_eta} (see Appendix \ref{ap_efficiency}). This
can be applied to the states found from the system's evolution via Hamiltonians Eqs.~\eqref{Hexp} and~\eqref{Hcom}, yielding the behavior depicted in  Fig.~\ref{fig:efficiency},  where the dotted curves coincide with the efficiency found from the conventional definition Eq.~\eqref{eq_eta} (continuous curves) under the same conditions.}

\change{The inset of  Figure~\ref{fig:efficiency}(a) plots the efficiency behavior under complete thermalization without dephasing. For small amplitude damping values, the efficiency remains below the Otto cycle efficiency limit. However, increasing the amplitude damping value allows the limit to be surpassed for  short periods of  time. On the other hand, the main Fig.~\ref{fig:efficiency}(a) shows the effect of amplitude damping on efficiency without dephasing and with $\lambda=0.5$, which produces a higher efficiency than complete thermalization. This result is a consequence of the increased work extraction shown in Figure~\ref{fig:RegimeHeatEngine}(a), where the system out of thermal equilibrium has an advantage in increasing the amplitude damping effect due to the rapid consumption of coherence.  
 This brings the system to a state similar to complete thermalization, as reflected in the temporal correlations  shown in Figure~\ref{fig:correlations}(b), which exhibits an exponential decay of correlations above unity due to quantum decoherence. In Figures~\ref{fig:correlations}(a) and \ref{fig:correlations}(c), partial thermalization and the absence of dephasing maintain temporal correlations for longer, due to the slow consumption of coherences.}

\change{Figure~\ref{fig:efficiency}(b) (inset) shows that dephasing with full thermalization does not generate an efficiency advantage, as the efficiency never exceeds the Otto efficiency limit. 
However, Figure~\ref{fig:efficiency}(b) (main) shows that under partial thermalization, the Otto limit can be slightly surpassed.  Increasing the phase damping parameter improves efficiency compared to low values of this parameter due to quantum coherence consumption.
Figure~\ref{fig:efficiency}(c), on the other hand, shows the combined effect of noise channels under partial thermalization ($\lambda=0.5$). In the main graph, with a fixed damping amplitude value $\gamma=0.8$, the efficiency exceeds the Otto limit. However, as dephasing increases, this advantage decreases. Meanwhile, the inset in Figure \ref{fig:efficiency}(c) shows that for a small damping amplitude value ($\gamma = 0.1$), the efficiency does not exceed the Otto limit. This confirms that the efficiency advantage is primarily produced by the damping amplitude channel.
}

\begin{figure*}[t]
    \centering
\includegraphics[width=0.9\linewidth]{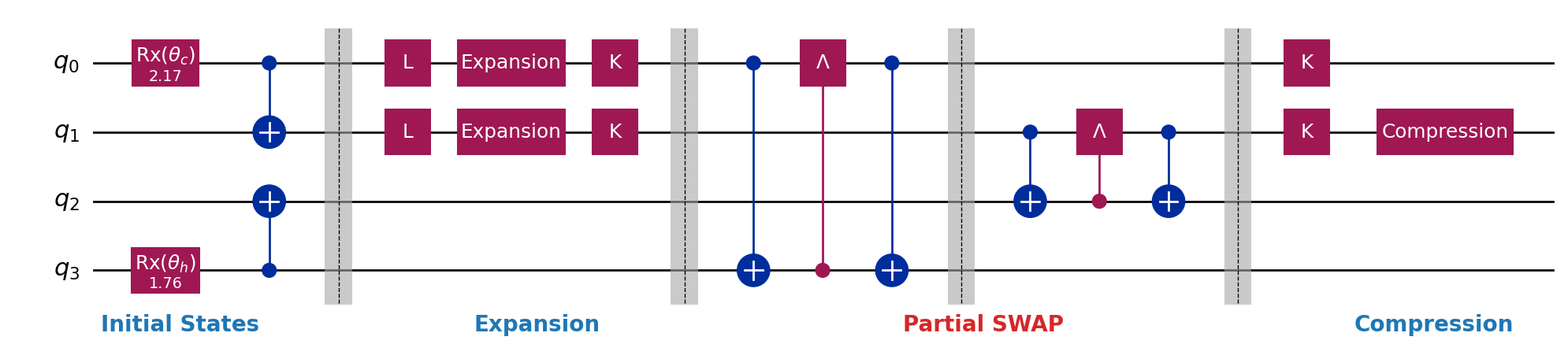}
    \caption{Quantum circuit implemented in \textit{Qiskit} \cite{qiskit2024}  to determine the work 
 extracted during the Otto cycle. Qubits $q_0$ and $q_3$ are ancillae to simulate the interaction with the reservoirs, $q_2$ is another ancilla qubit that helps to obtain the desired state in the partial SWAP, and qubit $q_1$ is the target from which the final states are obtained to find the expectation values of the energy.}
\label{fig:enter-label}
\end{figure*}



\change{The results for efficiency presented in Fig.~\ref{fig:efficiency} are always equal to or less than the Carnot limit, whether with full thermalization or partial SWAP, with or without noise. 
This behavior occurs because the proposed one-qubit heat engine cannot consume additional quantum resources, such as spatial correlations, to overcome the Carnot limit.} Therefore, in Eq.~\eqref{eq_efficiency}, the term due to entropy production from coherences must have a positive sign in the operating regime of the system as a heat engine, causing it to always subtract from the Carnot limit. \change{This connection between entropy production and the non-classical features of the dynamics has recently been explored~\cite{Upadhyaya2024}.}

\section{Simulating the cycle in a quantum circuit}
\change{To analyze the Otto cycle under realistic quantum hardware conditions, we introduce amplitude and phase damping channels. While these channels were previously presented  as part of the system's open dynamics, we focus here on simulating the typical noise processes found in commercial quantum devices, such as IBM Quantum processors. Amplitude damping represents energy relaxation and phase damping accounts for dephasing,
both of which are
common in near-term quantum processors~\cite{QCdiscovery,Kandala2017,Preskill2018}. Including these effects in our circuit simulation allows us to estimate the thermodynamic cost of implementing the engine on physical quantum platforms, as well as assessing the robustness of its quantum features in the presence of noise.} 

The proposed quantum engine was implemented in a quantum circuit using unitary gates, as shown in Fig.~\ref{fig:enter-label}. Three ancilla qubits and one target qubit were used.  We applied the  rotation gate $R_X$($\theta_i$), $i=c,h$, to two ancilla qubits to simulate thermalization with the cold and hot reservoirs, with rotation angles
\begin{equation}
\theta_i=2\arccos(\sqrt{p_i}),
\end{equation}
where \change{$p_i=\exp(-\beta_i\epsilon_i)/\mathcal{Z}_i$,  with $\epsilon_i$ the energy of the eigenstate $i$ and $\mathcal{Z}_i$ the partition function}. 
The expansion and compression gates were implemented from matrix representations of the unitary time evolution $\exp\left(-\frac{i}{h}\int_0^{\tau}\mathcal{H}_{k}dt\right)$, 
and the partial SWAP with the hot reservoir was implemented using two CNOT gates and \change{a controlled-gate c-$\Lambda\equiv I\otimes\left|0\right>\left< 0\right|+\Lambda\otimes\left|1\right>\left< 1\right|$, where} $\Lambda=\mathrm{I}\sqrt{1-\lambda}-\sigma_y\sqrt{\lambda}$, which is also applied to \textit{ancillae} qubits to obtain a copy of the initial states. \change{Additionally, the gates $L\equiv(\sigma_x+\sigma_z)/\sqrt{2}$ and $K\equiv(\sigma_y+\sigma_z)/\sqrt{2}$  were used to change the basis from computational to Pauli $X$ and $Y$, respectively.}

The described circuit  was simulated under different conditions. First, under closed conditions (without noise) to verify that the implemented circuit is consistent  with the unitary evolution model. Then, simulations were performed for different  amplitude and phase damping channels, and also for the noise associated with the implementation of the actual quantum logic gates, such as measurement errors and those associated with the Pauli representation.
\begin{figure}[b]
    \centering
\includegraphics[width=0.9\linewidth]{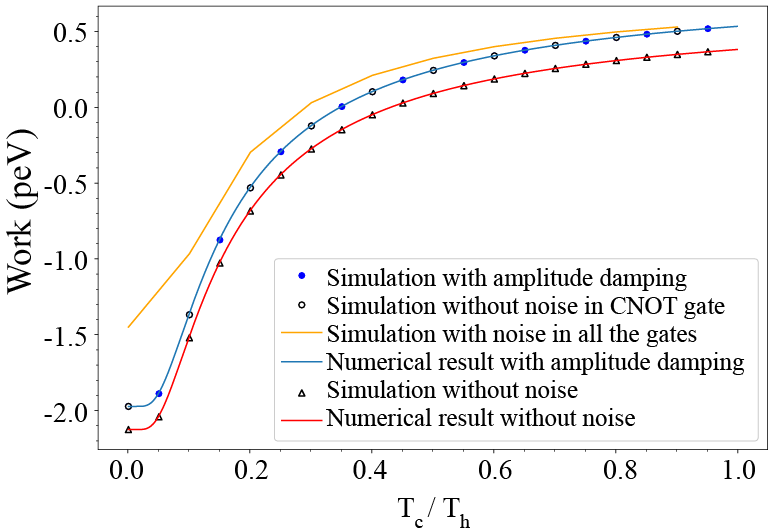}
\caption{Comparison of  simulations in a quantum circuit under noisy conditions with analytical results for the extracted work. \change{Simulations were performed with} partial thermalization $\lambda=0.6$, and noise  parameters $\gamma=p=0.6$. \change{$\omega_c$ and $\omega_h$ are the same as in Fig.~\ref{fig:Coherences}.}}
\label{fig:comparison_models}
\end{figure}

Figure~\ref{fig:comparison_models} shows that the quantum simulation results under 
unitary conditions agree with the obtained \change{numerical} results without noise, \change{ $\gamma=0$, $p=0$,}
(see unfilled triangles and continuous red curves in Fig.~\ref{fig:comparison_models}), so it can be concluded that the implemented circuit correctly represents the proposed model.  Figure~\ref{fig:comparison_models} also shows that the simulation under amplitude damping agrees with the \change{numerical} results for the same noise source~(see the blue dots and continuous blue curve), as well as with the simulation results when noise is applied to all gates in the circuit except the CNOT gate~(see unfilled circles). \change{However, when noise is added to the CNOT gate (shown by the continuous yellow curve), the simulation deviates from the numerical results. Thus, the  CNOT gate is the most} sensitive noise source in the circuit implementation.
\begin{figure}[t]
    \centering
\includegraphics[width=0.9\linewidth]{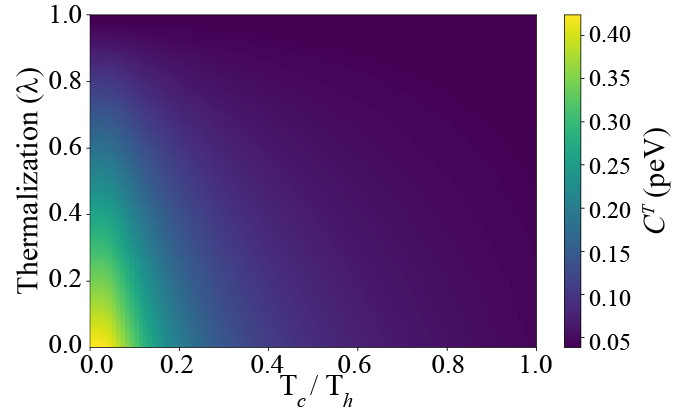}
\caption{Phase diagram of the thermodynamic cost $C^T$  as a function of the thermalization $\lambda$ and the temperature ratio $T_c/T_h$ in the reservoirs; \textcolor{black}{$\gamma=0.6$, $p=0.1$. $\omega_c$ and $\omega_h$ are the same as in Fig.~\ref{fig:Coherences}.}}
\label{fig:therm-cost}
\end{figure}

\change{\subsection{Thermodynamic Cost}}
We define a thermodynamic cost $C^T$ for  the performance of the quantum circuit implementing the heat engine under realistic noise conditions
as the difference between $\av{W}$  and that for the quantum simulations with noise in all applied gates (continuous yellow curve in Fig.~\ref{fig:comparison_models}), \change{say $\av{W_{\text{circ}}}$.  Thus, $C^T\equiv \av{W}-\av{W_{\text{circ}}}< \Delta_{\av{W}}$, where $\Delta_{\av{W}}\equiv \av{W}-\av{W_{\text{exact}}}$ with $\av{W}-\av{W_{\text{exact}}}$ being the work obtained from the numerical result without noise (see continuous red curve in Fig.~\ref{fig:comparison_models}).} 

As shown in Fig.~\ref{fig:therm-cost}, the thermodynamic cost depends not only on $\gamma$, which indicates the amount of energy dissipated, but also on the degree of thermalization and the temperatures in the reservoirs. 
Partial thermalization (small values of $\lambda$) and \change{small} temperature ratios generate a higher $C^T$. Since the thermodynamic cost is the difference between the \change{numerically calculated work with amplitude damping and the work obtained in the noisy circuit, it can be deduced that $C^T$ is mainly generated by noise in the CNOT gate. When the cycle is simulated without noise in the CNOT  gate, the circuit's work coincides with the numerically}
calculated work, as shown in Fig.~\ref{fig:comparison_models}.
\change{Appendix~\ref{ap_sigmaZ} illustrates a special case in which the Hamiltonians in the $\sigma_z$ basis allow us to obtain an analytical solution for the upper bound of $C^T$, which depends on the thermalization and decay parameters.  The thermodynamic analysis is based solely on the prescribed unitary strokes and dissipative thermalization processes.  Although the considered noise is relevant for accurately implementing experiments on quantum hardware, it does not modify the theoretical framework underlying this manuscript, which is hardware-independent.}

\section{Conclusions}
We performed a thermodynamic analysis of a one-qubit heat engine based on the quantum Otto cycle, incorporating partial thermalization and quantum noise. Our findings demonstrate that efficiency gains can be achieved by consuming coherence as a quantum resource. This improvement over the Otto limit persists even when using noisy states that are generated by amplitude and phase damping.

The study of the temporal correlations in the system revealed its quantum nature, \change{which is evidenced by coherences generated during expansion and compression processes}. This is quantified by the violation of the Leggett-Garg inequality, which decreases with amplitude and phase damping.  \change{The amount of extractable work is optimized when the system is fully thermalized or when dephasing is increased. Increasing the amplitude damping channel expands} the engine's operating regime and  \change{restricts the regime in which the system acts as a refrigerator}. In contrast, phase damping \change{increases the extractable work, but it doesn't offer any advantage in efficiency over the Otto limit.}

Efficiency \change{is limited by the Otto bound under full thermalization without noise. However,
it exceeds the bound  under partial thermalization, where quantum coherence is consumed in noisy conditions and correlates with enhanced performance in the presence of Leggett-Garg violations.}

Simulating the model with a quantum circuit under realistic noise conditions confirmed the accuracy of the implementation of unitary gates. The CNOT gate was identified as the main source of error, enabling the thermodynamic cost to be quantified under noisy conditions.

\begin{acknowledgments}
We thank the Norwegian Partnership Program for Global Academic Cooperation (NORPART), Grant NORPART-2021/10436 (QTECNOS).  
J.H.R. thanks the Department of Physics at the University of Oslo for their hospitality during his extended research stay. 
\end{acknowledgments}

\appendix

\section{\label{ap_dynamicMapping}Dynamic mappings}
The operators $E_i^{th}(\lambda)\equiv E_i^{tc}(\lambda)$ used in the mappings of strokes 2 and 4, $\rho_{\rm th}=\mathcal{E}_{\rm th}(\rho)$ and $\rho_{\rm tc}=\mathcal{E}_{\rm tc}(\rho)$ respectively, are given by
\begin{align}
    E_0^{th}&=
\sqrt{p_0}\begin{pmatrix}
        1 & 0 \\
        0 & \sqrt{1-\lambda}
    \end{pmatrix} , \,
        E_1^{th}=
\sqrt{p_0}\begin{pmatrix}
        0 & \sqrt{\lambda}\\
        0 & 0 
\end{pmatrix},\\  
E_2^{th}&=\sqrt{1-p_0}\begin{pmatrix}
        \sqrt{1-\lambda} & 0\\
        \nonumber
        0 & 1 
\end{pmatrix}, \,
       E_3^{th}= \sqrt{1-p_0}\begin{pmatrix}
        0 & 0\\
\sqrt{\lambda} & 0    \end{pmatrix},
\end{align}
where $p_0 =\exp({-\epsilon_h\beta_h})/\mathcal{Z}_h$ is the population of the system's excited state.

\change{
\section{\label{ap_masterEQ}Solution of the master equation during thermalization with a heat reservoir
}
The evolution of the open system during thermalization with the heat reservoir can be described using the master equation
\begin{equation}
    \frac{\partial \rho}{\partial t}=-\frac{i}{\hbar}[\mathcal{H},\rho]+\mathcal{D}_{\rm ad}(\rho),
    \label{meq}
\end{equation}
with dissipator $\mathcal{D}_{\rm ad}(\rho)=\sum_k[L_k\rho L_k^{\dagger}-\frac{1}{2}\{L_k^{\dagger}L_k,\rho\}]$. 
To solve this equation, we separate the time-dependent part of the Hamiltonian as follows: $\mathcal{H}(t)=\mathcal{H}_0+\mathcal{H}_I(t)$~\cite{breuer2002theory}. By rewriting Eq.~\eqref{meq} in the interaction picture, we obtain 
\begin{equation}
\frac{\partial\Tilde{\rho}(t)}{\partial t}=-\frac{i}{\hbar}\left[\mathcal{H}_I(t),\Tilde{\rho}(t)\right]+\gamma[\sigma_{-}\Tilde{\rho}(t)\sigma_{+}-\frac{1}{2}\{ \sigma_{+}\sigma_{-},\Tilde{\rho}(t)\}]
\end{equation}
where 
$\Tilde{\rho}(t)=e^{\frac{i}{h}\mathcal{H}_0t}\rho(t)e^{-\frac{i}{h}\mathcal{H}_0 t}$, $\gamma$ is the decay rate, and $\sigma_\pm$ are the raising and lowering operators.
Within the Born-Markov approximation we obtain~\cite{breuer2002theory}
\begin{equation}
\frac{\partial\Tilde{\rho}(t)}{\partial t}=\gamma[\sigma_{-}\Tilde{\rho}(t)\sigma_{+}-\frac{1}{2}\{ \sigma_{+}\sigma_{-},\Tilde{\rho}(t)\}].
\end{equation}
In terms of the  Bloch vector representation, this equation has the well-known solution 
\begin{equation}
    \Tilde{\rho}(t)=\frac{1}{2}(\mathbb{I}+\left< \Vec{\sigma}(t)\right>\cdot\Vec{\sigma})= \begin{pmatrix}
        \frac{1}{2}(1+\left<\sigma_z(t)\right>) & \left< \sigma_{-}(t)\right>\\
        \left< \sigma_{+}(t)\right> &  \frac{1}{2}(1-\left<\sigma_z(t)\right>)
    \end{pmatrix},
\end{equation}
from which we obtain the expectation values:
\begin{align}
    \left< \sigma_x(t)\right>=&\left<\sigma_x(0)\right> e^{-\frac{\gamma}{4}t},\\
\left<\sigma_y(t)\right>=&\left<\sigma_y(0)\right> e^{-\frac{\gamma}{4}t},\\
\left<\sigma_z(t)\right>=&\left<\sigma_z(0)\right> e^{-\frac{\gamma}{2} t}+1-e^{-\frac{\gamma}{2} t}.
\end{align}
Setting $\gamma'=1-e^{-\frac{\gamma}{2} t}$ yields a solution  equivalent to the dynamical mapping $\Tilde{\rho}(t)=\mathcal{E}(\Tilde{\rho}(0))=\Xi_0\Tilde{\rho}(0)\Xi_0^\dagger+\Xi_1\Tilde{\rho}(0)\Xi_1^\dagger$, with \textit{amplitude} damping via the Krauss operators
\begin{equation}
    \Xi_0=
    \begin{pmatrix}
        1 & 0 \\
        0 & \sqrt{1-\gamma'}
    \end{pmatrix}, \;  \text{and} \;\; 
        \Xi_1=
    \begin{pmatrix}
        0 & \sqrt{\gamma'}\\
        0 & 0 
    \end{pmatrix}. 
\end{equation}
}

\change{Similarly, to account for the phase damping effect (Eq.~\eqref{eq:krausPD}), we assume that the energy levels remain constant and the relative phase between the states $|0\rangle$ and $|1\rangle$ is decaying. This process can be represented by the master equation 
$  \frac{d\rho}{d t}=-i[\mathcal{H},\rho]+\mathcal{D}_{\rm pd}(\rho)$, 
where $\mathcal{D}_{\rm pd}(\rho)=\Gamma(\sigma_z\rho\sigma_z-\{\sigma_z^\dagger\sigma_z; \rho\})$, 
 and $\Gamma$ is the dephasing rate.
 Rewriting the equation in the interaction picture, within the Born-Markov approximation 
 yields the following differential equations for the density matrix elements:
$\dot{\rho}_{00}=\dot{\rho}_{11}=0$, $\dot{\rho}_{01}=-2\Gamma\rho_{01}$, $\dot{\rho}_{10}=-2\Gamma\rho_{10}$.
 Integrating respect to time, we obtain
 \begin{equation}
    \label{eq:solutionPD}
     \rho(t)=
     \begin{pmatrix}
         \rho_{00} & e^{-2\Gamma t}\rho_{01}\\
        e^{-2\Gamma t}\rho_{10} & \rho_{11} 
     \end{pmatrix}.
 \end{equation}
 Setting $1-2p\equiv e^{-2\Gamma t}$ makes the solution to Eq.~\eqref{eq:solutionPD} equivalent to the dynamical mapping
\begin{equation}
    E_0^p=\begin{pmatrix}
        1 & 0\\
        0 & \sqrt{1-p}
    \end{pmatrix} ,\hspace{1cm}
    E_1^p=\begin{pmatrix}
        0 & 0\\
        0 & \sqrt{p}   
    \end{pmatrix}.
\end{equation}
If we consider the effects of amplitude and phase damping channels, the master equation becomes
\begin{equation}
    \frac{d\rho}{d t}=-i[\mathcal{H},\rho]+\mathcal{D}_{ad}(\rho)+\mathcal{D}_{pd}(\rho),
\end{equation}
whose elements, in the interaction picture and within the Born-Markov approximation, are given by 
\begin{align}
    \dot{\rho}_{11}&=-\gamma\rho_{11},\\
    \dot{\rho}_{01}&=-\left(\frac{\gamma}{2}+2\Gamma\right)\rho_{01}\\
    \dot{\rho}_{10}&=-\left(\frac{\gamma}{2}+2\Gamma\right)\rho_{10}.
\end{align}
This leads to the solution
\begin{equation}
    \label{eq:combinedAdPd}
    \rho(t)=
    \begin{pmatrix}
    1-\rho_{11}(0)e^{-\frac{\gamma}{2}t} & \rho_{01}(0)e^{-(\frac{\gamma}{2}+2\Gamma)t}\\
    \rho_{10}(0)e^{-(\frac{\gamma}{2}+2\Gamma)t} & \rho_{11}(0)e^{-\frac{\gamma}{2}t}
    \end{pmatrix}.
\end{equation}
Equation~\eqref{eq:combinedAdPd} is equivalent to applying the amplitude and phase damping 
dynamic mappings independently, such as 
with the operator $\rho(t)=\mathcal{E_{\rm ad}(\mathcal{E}_{\rm pd}(\rho(\rm 0 )))}$.
 }

\change{
\section{\label{ap_sigmaZ}Alternative quantum heat engine}

In an alternative, complementary model, the Hamiltonians in equations~\eqref{Hexp} and ~\eqref{Hcom} can be substituted by their counterparts in the Pauli $\sigma_z$ basis, thus obtaining $H_\text{exp}(t)=-\frac{1}{2}\hbar\omega(t)\sigma_z$, and $H_\text{comp}(t)=-\frac{1}{2}\omega(\tau-t)\sigma_z$.
}
\textcolor{black}{
In this scenario, the unitary time evolution of the system admits an analytical solution for the extractable work:
\begin{equation}
\av{W_{\text{exact}}}=\frac{\Delta\epsilon}{2} \lambda \left[\tanh \left(\frac{\beta_c\epsilon_c}{2}\right)-\tanh \left(\frac{\beta_h \epsilon_h}{2}\right)\right],
\label{eq_exact_work}
\end{equation}
where $\Delta\epsilon=\epsilon_c - \epsilon_h$. On the other hand, the calculation for the work expectation value under amplitude and phase damping gives
\begin{equation}
    \av{W}= {\text{e}}^{-\frac{\gamma t}{2}}\av{W_{\text{exact}}}-(1-{\text{e}}^{-\frac{\gamma t}{2}})\av{E_{\text{dis}}},
    \label{eq_work}
\end{equation}
where
$\av{E_{\text{dis}}}=\frac{\Delta\epsilon}{2}\left[1-\tanh\left(\frac{\beta_c \epsilon_c}{2}\right)\right]$
gives the energy dissipated by the system's interaction with the environment.
From these equations, it can be seen that, in the absence of thermalization ($\lambda=0$), the exact extracted work is zero. Increasing the transition rate $\gamma$ causes the expectation value of the work to decay exponentially. When this decay parameter vanishes, the extracted work corresponds to the exact result. The phase damping parameter does not appear in the equations since this form of noise causes decoherence without energy loss~\cite{schlosshauer2019quantum, nielsen2001quantum}.}

\textcolor{black}{For the sake of completeness, an expression can be derived for the thermodynamic cost, which yields an upper bound:
\begin{equation}
   C^T\equiv \av{W}-\av{W_{\text{circ}}}< \Delta_{\av{W}} ,
\end{equation}
$\Delta_{\av{W}}\equiv \av{W}-\av{W_{\text{exact}}}=({\text{e}}^{-\frac{\gamma t}{2}}-1)(\av{W_{\text{exact}}}+\av{E_{\text{dis}}})$. $C^T$ is upper bounded by $\Delta_{\av{W}}\equiv\frac{\Delta\epsilon}{2}\left({\text{e}}^{-\frac{\gamma t}{2}}-1\right)\times\left[1+(\lambda-1)\tanh\left(\frac{\beta_c\epsilon_c}{2}\right)-\lambda\tanh\left(\frac{\beta_h  \epsilon_h}{2}\right)\right]$.
}

\textcolor{black}{Leggett-Garg temporal correlations also admit an analytical solution:
\begin{eqnarray}
       K & = & \text{sech}(\phi)\Big(\cos(\Omega_ht+i\phi)+\sqrt{1-\lambda }\sqrt{1-p} \nonumber \\ & & 
 \times\sqrt{{\rm e}^{- \gamma  t}}\left[\cos(\Omega_ct+i\phi)-\cos \left(\Omega t+i\phi\right)\right]\Big),
 \label{eq_correlations}
\end{eqnarray}
which explains its oscillatory behavior. Here, $\Omega=\omega_c+\omega_h$, $\Omega_h=\omega(\tau-t)+\omega_h$, $\Omega_c=\omega(t)+\omega_c$, and  $\phi=\frac{\beta_c\epsilon_c}{2}$.  
The model's efficiency in the $\sigma_z$ basis is consistent with the quantum Otto cycle limit.}

\textcolor{black}{\section{\label{ap_efficiency}Calculating efficiency as relative entropy}
We apply the relative entropy $D[\rho||\sigma]\equiv\text{Tr}[\rho\text{ln}\rho]-\text{Tr}[\rho\text{ln}\sigma]$ to calculate the entropy production during the cycle:
\begin{equation}  S(\rho||\sigma)=D\left[\rho_\text{exp}(\tau)||\rho_\text{th}^f \right]+ D\left[\rho_\text{comp}(\tau)||\rho_{c}^f \right].
\end{equation}
Since von Neumann's entropy  $S(\rho) = \mathrm{Tr}(\rho \ln \rho)$ is invariant under unitary evolution, $S(\rho(\tau)) = S(\rho(0))$, the expansion and compression processes can be written as $\mathrm{Tr}[\rho_\text{exp}(\tau) \ln \rho_\text{exp}(\tau)] = \mathrm{Tr}[\rho_c \ln \rho_c]$, and $\mathrm{Tr}[\rho_\text{comp}(\tau) \ln \rho_\text{comp}(\tau)] = \mathrm{Tr}(\rho_\text{th}^{\mathrm{f}} \ln \rho_\text{th}^{\mathrm{f}})$. We thus  obtain
\begin{align}
S(\rho||\sigma)&= \mathrm{Tr} \left[ \left( \rho_c - \rho_\text{comp}(\tau) \right) \ln \rho_c \right] \\ \nonumber
   &+\mathrm{Tr} \left[ \left( \rho_\text{th}^f - \rho_\text{exp}(\tau) \right) \ln \rho_\text{th}^f \right]. 
\end{align}
Using the thermal state $\rho_c = \frac{e^{-\beta_c \mathcal{H}_c}}{\mathcal{Z}_c}$ and defining the heat from the cold reservoir as $\av{Q_c}=\mathrm{Tr} \left[ (\rho_c - \rho_\text{comp}) \mathcal{H}_c \right]$, we write the entropy production as
\begin{equation}
S(\rho||\sigma) = -\beta_{\mathrm{c}} \langle Q_{\mathrm{c}} \rangle +\mathrm{Tr} \left[ \left( \rho_\text{th}^f - \rho_\text{exp}(\tau) \right) \ln \rho_\text{th}^f \right] .
\end{equation}
Thus,  taking into account the lost energy $\Delta E$, the first law reads 
$\langle Q_{\mathrm{c}} \rangle + \langle Q_{\mathrm{h}} \rangle + \langle W \rangle = 0$, and we have
\begin{align}
S(\rho||\sigma) &= \beta_{\mathrm{c}} \langle W \rangle + \beta_c\av{Q_h}\\ \nonumber
&+\mathrm{Tr} \left[ \left( \rho_\text{th}^f - \rho_\text{exp}(\tau) \right) \ln \rho_\text{th}^f \right] .
\end{align}
Rearranging, we  obtain
\begin{equation}
\label{eq:eff_new}
- \frac{\langle W \rangle}{\langle Q_{\mathrm{h}} \rangle} = 1+\frac{\mathrm{Tr} \left[ \left( \rho_\text{th}^f - \rho_\text{exp}(\tau) \right) \ln \rho_\text{th}^f \right] }{\beta_c \av{Q_h}} -\frac{S(\rho||\sigma)}{\beta_c \av{Q_h}}.
\end{equation}
Adding and substracting the term $\beta_c\av{Q_h}$ to 
$\mathrm{Tr} \left[ \left( \rho_\text{th}^f - \rho_\text{exp}(\tau) \right) \ln \rho_\text{th}^f \right]
$,
and since
$\av{Q_h}\equiv\mathrm{Tr}[(\rho_\text{th}^f-\rho_\text{exp})\mathcal{H}_h]$, 
\begin{align}
   -\beta_h\av{Q_h}&= \mathrm{Tr}[(\rho_\text{th}^f-\rho_\text{exp})\ln(e^{-\beta_h\mathcal{H}_h}/\mathcal{Z}_h)]\\ \nonumber
    &=\mathrm{Tr}[(\rho_\text{th}^f-\rho_\text{exp})\ln(\rho_h)],
\end{align}
where we have taken into account that $\mathrm{Tr}[(\rho_\text{th}^f-\rho_\text{exp})\mathcal{Z}_h]=0$. Hence,
\begin{align}
 &\mathrm{Tr} \left[ \left( \rho_\text{th}^f - \rho_\text{exp}(\tau) \right) \ln \rho_\text{th}^f \right]=\\ \nonumber
    &\mathrm{Tr} \left[ \left( \rho_\text{th}^f - \rho_\text{exp}(\tau) \right) \ln \frac{\rho_\text{th}^f}{\rho_h}\right]-\beta_c\av{Q_h}.
\end{align}
Substituting this result into Eq.~\eqref{eq:eff_new}, $\eta$ becomes
\begin{equation}
\eta=-\frac{\av{W}}{\av{Q_h}}=\eta_{Carnot}-\frac{\Delta S_\text{th}}{\beta_c\av{Q_h}}\equiv \eta_E,
\end{equation}
where $\eta_{Carnot}=1-\frac{\beta_h}{\beta_c}$ and $\Delta S_{th}=S(\rho||\sigma)-\mathrm{Tr} \left[ \left( \rho_\text{th}^f - \rho_\text{exp}(\tau) \right) \ln \frac{\rho_\text{th}^f}{\rho_h}\right].$
}

\nocite{*}
\bibliography{Bibliography}

\end{document}